\documentclass[aps,pre,twocolumn,showpacs]{revtex4}

\usepackage{graphicx}
\usepackage{amssymb}
\usepackage{amsmath}

\newcommand{\ie}{{\it i.e.\/}}
\newcommand{\avem}{\langle m \rangle}
\newcommand{\avel}{\langle \lambda \rangle}

\begin{document}

\title{Coarsening of Sand Ripples in Mass Transfer Models with
Extinction}  

\author{E. K. O. Hell\'en}
\email{ehe@fyslab.hut.fi}
\affiliation{Laboratory of Physics, Helsinki University of Technology,
P. O. Box 1100, FIN-02150 HUT, Finland}
\author{J. Krug}
\email{jkrug@Theo-Phys.Uni-Essen.DE}
\affiliation{Fachbereich Physik, Universit\"at Essen, 45117 Essen, Germany}

\date{\today}

\begin{abstract}
Coarsening of sand ripples is studied in a one-dimensional stochastic
model, where neighboring ripples exchange mass with algebraic rates, 
$\Gamma(m) \sim m^\gamma$, and ripples of zero mass are
removed from the system. For $\gamma < 0$ ripples vanish through
rare fluctuations and the average ripples mass grows as
$\avem(t) \sim -\gamma^{-1} \ln (t)$. Temporal correlations decay
as $t^{-1/2}$ or $t^{-2/3}$ depending on the symmetry of the 
mass transfer, and asymptotically the system is
characterized by a product measure. The stationary ripple mass distribution
is obtained exactly. For $\gamma > 0$ ripple evolution is linearly
unstable, and the noise in the dynamics is
irrelevant. For $\gamma = 1$ the problem is solved
on the mean field level, but the mean-field theory does not
adequately describe the full behavior of the coarsening. In particular,
it fails to account for the numerically observed universality with respect
to the initial ripple size distribution.
The results are not restricted to sand ripple
evolution since the model can be mapped to zero range processes,
urn models, exclusion processes, and cluster-cluster aggregation.

\end{abstract}

\pacs{45.70.Qj, 47.54.+r, 05.45.-a, 05.10.-a}

\maketitle


\section{Introduction and Motivation}
\label{Introduction}

When a surface of sand is exposed to wind or water flow, 
patterns like ripples or dunes are commonly
formed. The physics of this process is extremely complex
because it involves the interaction of a granular medium 
with a possibly turbulent hydrodynamic flow
\cite{Bagnold}. It is therefore desirable to develop
simplified models that capture some of the key features of the pattern
formation. 

\begin{figure}
\centering
\includegraphics[width=\linewidth
]{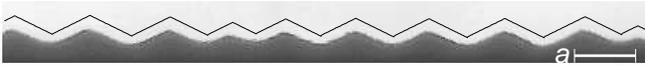}
\caption{Experimental image of vortex ripples in a one-dimensional annular
geometry~\cite{Andersen:preprint}. The amplitude of the fluid
oscillations is denoted by $a$. The line above the pattern shows a fit
of triangles with a constant slope. Courtesy of K.H.~Andersen.  
}
\label{Vortex_ripples}
\end{figure}

In this paper we are concerned with a class of models which focus
on the role of the mass transfer in the evolution of the
pattern. Along a one-dimensional
cut perpendicular to the ripples, the pattern is described
by a set $\{\lambda_i \}$ of ripple lengths, where the index
$i$ labels the ripples in the array. The $\lambda_i$ are used
here as a general measure of ripple size, without reference to the detailed
geometry of individual ripples (see Fig.\ref{Vortex_ripples}).
In particular, we do not distinguish between the linear size of a ripple
and the mass it contains (for further discussion of this point
see~\cite{Andersen:PRE63}). 

During the evolution of the patterns, the flow transfers mass between
neighboring ripples. The central assumption of the model is that the
mass transferred to ripple $i$ from ripple $i+1$ or $i-1$ (per unit time) 
is a function $\Gamma(\lambda_i)$ of the size of the
ripple which gains the mass. Further motivation for this
assumption will be given below. We refer
to $\Gamma(\lambda)$ as the \emph{robber function}~\footnote{This term
was  suggested to us by Ko van der Weele.}.

Depending on the characteristics of the flow, the mass transfer
between ripples can be symmetric or asymmetric. In the symmetric
case the balance between loss and gain processes for a given
ripple leads to the evolution equation~\cite{Andersen:PRE63}
\begin{equation}
\frac{{\rm d}\lambda_i}{{\rm d}t} = \frac{1}{2}[
-\Gamma(\lambda_{i-1}) + 2 \Gamma(\lambda_i) - \Gamma(\lambda_{i+1})
], \label{detlambdaeq_sym}
\end{equation}
while in the asymmetric case (assuming, say, mass transfer only 
to the left) one has
\begin{equation}
\frac{{\rm d}\lambda_i}{{\rm d}t} = -\Gamma(\lambda_{i-1}) +
\Gamma(\lambda_i).
\label{detlambdaeq_asym}
\end{equation}
The factor $1/2$ in Eq.~\eqref{detlambdaeq_sym} makes the time
scales for both the dynamics equal. 

A homogeneous state of equally sized ripples, $\lambda_i \equiv \bar \lambda$,
is stationary under \eqref{detlambdaeq_sym} and \eqref{detlambdaeq_asym},
but its stability depends on the derivative of the 
robber function: The pattern is stable for $\Gamma'(\bar \lambda) < 0$
and unstable for $\Gamma'(\bar \lambda) > 0$~\cite{Andersen:PRE63}.
In the unstable case the dominant mode is a modulation of period 2, in which
every second ripple grows and every second one shrinks. As the size of 
the shrinking ripples reaches zero in a finite time, the evolution 
equations~\eqref{detlambdaeq_sym} and \eqref{detlambdaeq_asym} have
to be supplemented by an \emph{extinction rule}: When the size of 
a ripple vanishes, it is removed from the system and the remaining
ripples are relabeled such that the previous neighbors of the removed
ripple become neighbors of each other. Extinction events contribute
to the \emph{coarsening} of the pattern, \ie, to an increase of the mean
wavelength. In this work the reverse process of ripple creation is not 
considered, hence coarsening is irreversible.  

The symmetric mass transfer model \eqref{detlambdaeq_sym} was first
proposed as a description of 
vortex ripples in coastal waters, which are created
under the oscillatory flow of surface waves~\cite{Andersen:PRE63}.
In that context the dependence of the robber function on the size of
the gaining ripple is motivated by the observation that the mass transfer
is effectuated mostly by a separation vortex which appears in the wake
of that ripple. Numerical simulations~\cite{Andersen:PRE63} and 
experiments~\cite{Andersen:preprint} show that $\Gamma(\lambda)$
is nonmonotonic, with a maximum near $\lambda = a$, where $a$ is 
the amplitude of the fluid oscillations. Thus patterns of wavelength
$\bar \lambda < a$ ($\bar \lambda > a$) are unstable (stable), and
the main interest is in the wavelength selection process starting from
a short wavelength, unstable state 
\cite{Andersen:PRE63,Andersen:preprint,Krug:ACS}.  

A related, asymmetric mass transfer model for wind driven sand
ripples was introduced in~\cite{Werner:PRL71}. 
The basic hypothesis of the model is that wind ripples wander with
a speed that is inversely proportional to their size.
This implies that a leading ripple (ripple $i+1$) is eroded by the
trailing ripple (ripple $i$) at a rate which is proportional to
$1/\lambda_{i}$, so the resulting evolution equation is of the
type~\eqref{detlambdaeq_asym} with $\Gamma(\lambda) \sim 1/\lambda$. 
Since $\Gamma'(\lambda) < 0$, the homogeneous pattern is stable.
However, when fluctuations are included by discretizing the
ripple sizes and implementing a stochastic
mass transfer rule,
a fluctuation-driven coarsening mechanism becomes effective and
leads to an increase of the mean wavelength with time $t$ as $\ln t$.

In this article we consider a class of stochastic models whose noiseless 
counterparts are described by~\eqref{detlambdaeq_sym} or
\eqref{detlambdaeq_asym}. We
concentrate on monotonic, algebraic robber functions $\Gamma(\lambda) \sim
\lambda^\gamma$ and study the coarsening process regarding $\gamma$ as
a variable parameter. For $\gamma < 0$ this extends the results of
\cite{Werner:PRL71} on fluctuation-driven coarsening.
The case $\gamma > 0$ is a simple realization of linearly
unstable ripple evolution, and it is studied here as a first step
towards a better understanding of models with nonmonotonic robber
functions~\cite{Andersen:PRE63,Andersen:preprint,Krug:ACS}. 
Although the models are
defined using the terminology of sand ripples, they are connected to
other problems in nonequilibrium statistical physics. 
For example, for $\gamma=0$ the system maps to
coalescing random walks and is therefore exactly solvable. Other equivalences
include exclusion processes, zero range processes,
urn models, and cluster-cluster aggregation. 

Our main results are the following. In general, one can
identify two time scales in the dynamics: 
The one of ripple extinctions and the other 
at which the system would equilibrate to a steady state
in the absence of extinctions. For $\gamma < 0$ the loss of a
ripple is a rare fluctuation when the mean ripple size is large. 
Therefore the two time scales are well separated, and
the system has 
time to relax to a quasi-steady state between ripple extinctions.
We show that this state is characterized by a product measure.
This justifies the mean field assumption made in~\cite{Werner:PRL71},
and allows us to calculate the stationary ripple size distribution. 
The product measure becomes exact only at the limit $t \to \infty$
as the correlations in the system decay as a power law. 
The average ripple
size grows logarithmically at late times, with a prefactor
$-\gamma^{-1}$. 

For $\gamma > 0$ extinctions are frequent events which occur on the same
time scale as the evolution of the surviving ripples. We find in this
case that the noise is irrelevant, so that the dynamics   
can be described by \eqref{detlambdaeq_sym} and \eqref{detlambdaeq_asym}.
For $0< \gamma < 1$ the mean ripple size grows 
algebraically with the exponent $1/(1-\gamma)$, while the growth
is exponential for $\gamma = 1$. In the latter case the evolution 
equations become linear, and the problem can be solved exactly on the
mean field level. The mean field theory  
reproduces the  exponential growth for the mean ripple size,
but incorrectly predicts a dependence of the
ripple size distribution and the coarsening law on the initial conditions. 

In the next section the model is introduced and its relations
to other models are discussed. Algebraically
decaying robber functions ($\gamma < 0$) are considered in
Sec.~\ref{admtfsec}. The product form of the 
mass distribution is derived in Sec.~\ref{massdistsec}, the
coarsening law is calculated in Sec.~\ref{coarssec}, and the approach
to the product measure is analyzed in Sec.~\ref{correlationssec}. Section~\ref{agmtfsec}
is devoted to algebraically growing robber functions
($\gamma > 0$). The
mean-field theory is first developed for $\gamma=1$
and then compared to simulations
(Sec.~\ref{MFgamma1sec}). Section~\ref{gbtw01sec} examines the case 
$0<\gamma<1$. Conclusions and open questions are formulated in
Section~\ref{conlsec}.

\section{The stochastic ripple model} \label{modelsection}

\subsection{Definition and simulation algorithm}

In the stochastic model a sand ripple is characterized by its
mass $m$. The mass variables are integers such that each ripple
consists of $m_i$ elementary mass units and occupies a site $i$ on
a one-dimensional lattice. The mass is conserved, \ie, M := $\sum_i m_i 
= {\rm   const.}$. The $m_i$
correspond to the length variables $\lambda_i$ used in
equations~\eqref{detlambdaeq_sym} and \eqref{detlambdaeq_asym}. As
mentioned in 
the Introduction, the mass and the length of ripples are here
considered to be indistinguishable. We use different symbols for two
reasons. We want to make a clear  
distinction between (i) the real and integer valued ripple sizes and
(ii) between the deterministic and noisy dynamics. 

Ripples interact only by exchanging mass with their nearest
neighbors with an algebraic mass transfer rates $\Gamma(m) =
\Gamma_0 m^\gamma$. Since the constant $\Gamma_0$
affects only the time scale it will be set equal to unity from now
on. If ripples obtain
mass only from one of their neighbors, say, from the right one, the
mass transfer is called (totally) asymmetric. If the mass comes from
both neighbors we call the dynamics symmetric. 
As was discussed in Section~\ref{Introduction}, the asymmetric mass
transfer naturally arises in the case of wind ripple
formation~\cite{Werner:PRL71} whereas the symmetric dynamics
takes place for ripple patterns 
forming under an oscillatory flow~\cite{Andersen:PRE63}. 

In addition the model includes the removal
of ripples when their mass becomes zero. This is done such that 
lattice sites containing no mass are eliminated from the system. In
this way each ripple always has a neighbor from which it can gain
mass. If we denote the number of lattice sites at time $t$ by
$N(t)$, the average ripple mass is $\avem(t) = M/N(t)$. 

In the simulations three different initial conditions are used. As
random initial conditions we denote the case in which the probability
to have a ripple of size $m$ is given by the geometric
distribution $(1-q)q^{m-1}$, with $0 < q < 1$. 
The probability $q$ is
related to the mean ripple size as $\avem = (1-q)^{-1}$. A
distribution $m_i = \avem \ \forall i$ is referred to as monodisperse. The
third possibility is a Poisson distribution. 

The dynamics is implemented as follows. First a ripple is selected
randomly and time is 
incremented by $N(t)^{-1}\Gamma_{\rm max}^{-1}$, where 
$\Gamma_{\rm max}$ is the maximum of all the rates of the ripples in the
system at time $t$. Denote the mass of the selected ripple by $m$. 
If $x<\Gamma(m)/\Gamma_{\rm max}$, where $x$ is an
uniformly distributed random number in the interval~$[0,1]$, the
ripple gets a unit mass from its nearest neighbor. Otherwise a new
ripple is selected and the process is repeated. For
symmetric dynamics the neighbor is selected randomly whereas in the
asymmetric case it is always the right one.

\subsection{Relation to other models} \label{reltootmod}

The model defined above is inspired by the worm
model originally introduced to describe the coarsening of
wind ripples~\cite{Werner:PRL71}. Here it is generalized
in two respects. First, the mass transfer rate in the worm model is
inversely proportional to the ripple mass whereas in the generalized
model it is given by $\Gamma(m) \sim m^\gamma$. Second, we consider
also the symmetric mass transfer between neighbors. In the case of
asymmetric dynamics and for $\gamma = - 1$ our model reduces to the
worm model.

Apart from the extinction step, the sand ripple model is similar
to a zero range process~\cite{Spitzer:AIM5,Spohninkirja,Evans:BJP}.
Both models are defined in terms of conserved, integer mass variables
$m_i$ which interact through the (symmetric or asymmetric) exchange of 
unit masses between nearest neighbor sites of a lattice. 
The key difference is that in a zero range
process the mass transfer rate is a function of the mass at the
site of \emph{departure}, while in the sand ripple model it depends on the
mass at the \emph{target} site. This reverses the sign of the right hand sides 
of (\ref{detlambdaeq_sym}) and (\ref{detlambdaeq_asym}), and hence the
stability properties of the model: In a zero range process the homogeneous
state is stable if $\Gamma'(\bar m) > 0$ and unstable if
$\Gamma'(\bar m) < 0$. The coarsening behavior in zero range processes
with nonmonotonic robber function is relevant to 
clustering in granular gases~\cite{vanderMeer:PRE63}.

The occurrence of irreversible extinction events in our model is
reminiscent of certain urn models that have been proposed in the context
of glassy dynamics~\cite{Godreche:}. 
For example, consider the backgammon
model~\cite{Ritort:PRL75} which is defined by $M$ particles
distributed among $N$ boxes with $m_i$ 
particles in the $i$th box. The Hamiltonian is $H = -\sum_i
\delta_{{m_i},0}$ so that the energy corresponds to minus the number
of empty boxes. The $N$-fold degenerate ground state therefore consists
of a condensate, where all the particles belong to one
box. Associating the masses of ripples with the particle numbers
and lattice sites
with boxes, the ripple evolution becomes 
similar to the backgammon model at zero temperature, where
the empty boxes are not refilled once they have become empty (ripple
extinction). 

In contrast to the ripple model,
the urn models have no spatial structure, \ie, mass transfer is possible
between any pair of boxes.
In the standard dynamical scheme, originally
due to Ehrenfest, in each time step one of
the balls is chosen at random and a move to another box is attempted
\cite{Godreche:}. The probability for a box to be
chosen is then proportional to its occupation number. In our
setting this corresponds to a mass transfer rate $\Gamma(m_i) \sim m_i$,
where $i$ is the site of \emph{departure}; in this respect the urn
models are related to zero range processes.
Since $\Gamma' > 0$, the homogeneous state is linearly stable
and coarsening (\ie, evolution towards the ground state) is very
slow.

\begin{figure}
\centering
\includegraphics[width=\linewidth
]{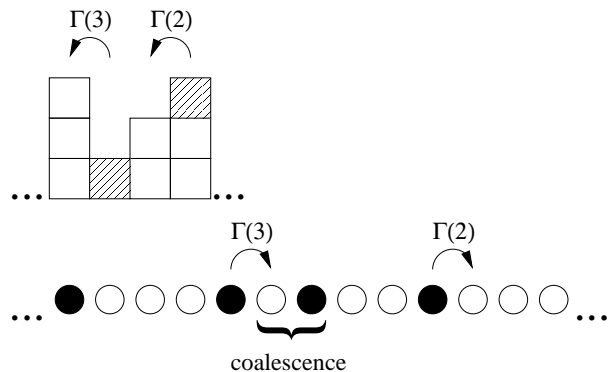}
\caption{Mapping between the asymmetric worm model and the exclusion
  process with coalescence.
}
\label{WORM_mappings}
\end{figure}

The sand ripple model can also be mapped to an exclusion
process~\cite{Spohninkirja,Liggetinkirja,DerridaEvans_in_Privmans_book}.
The mapping can be done in two ways which differ in how the disappearance
of ripples is taken into account. The mappings proceed along the
lines of~\cite{Majumdar:JSP99} and the first one is 
schematically presented in Fig.~\ref{WORM_mappings}. One constructs a 
new lattice with $L(t)=M+N(t)$ sites. The mass variables $m_i$ of the
ripples turn to $m_i$ consecutive holes  
separated by particles on the new lattice. More precisely, there
exist particles on sites $i+\sum_{k=1}^i m_k$
($i=1,\ldots,N(t)$) while the rest are empty. Moving one mass unit from one
ripple to its neighbor corresponds to a hop of a particle in the
exclusion process. 
Naturally the exclusion process is either symmetric or
asymmetric as the original dynamics.   
For the exclusion
process the loss of a ripple becomes a coalescence of particles at
contact, which changes the length $L(t)$ of the system.

As the masses map to holes, the hopping rates of
particles depend on the distances between them. For
negative (positive) $\gamma$ there is a repulsive (attractive)
interaction between the particles. For the marginal case of mass
independent transfer rates ($\gamma=0$) the interaction vanishes and
the exclusion process reduces to coalescing random 
walkers. This is a well-known problem which can be solved exactly
e.g. with the method of interparticle distribution
functions~\cite{ben-Avraham_in_Privmans_book}. The most relevant
results for our case are: (i) the average ripple mass grows
asymptotically as $\avem(t) \sim \sqrt{t}$ and (ii) the ripple size
distribution (the probability of finding a ripple of mass $m$ at time
$t$) can be written in a scaling form as $p(m;t) = m^{-1}
G(m/\avem)$, where the scaling function $G(x) = \frac{1}{2}\pi x^2
e^{-\pi x^2/4}$.

\begin{figure}
\centering
\includegraphics[width=\linewidth
]{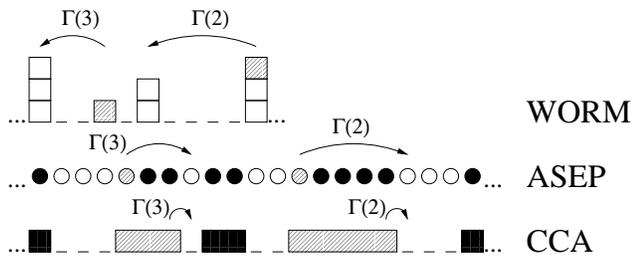}
\caption{Mapping between the worm model, the asymmetric exclusion
process (ASEP), and the cluster--cluster aggregation (CCA).
}
\label{WORM_DLCAmapping}
\end{figure}

A length conserving mapping is obtained by considering the model
with ripple extinction but without removal of empty lattice sites. 
Particles hop over these sites in order not to allow for creation
of new ripples. Again the empty sites become particles in the
exclusion picture but now the mass transfer corresponds to a hop of a
particle over all the particles in the same cluster. This is
illustrated in Fig.~\ref{WORM_DLCAmapping}. 

A long one particle hop can be considered as  
moving a cluster of particles as a whole. The loss of a ripple turns
into the aggregation of clusters. In this way our model further maps
to a cluster--cluster aggregation process where each cluster moves
with a rate that depends on the distance to its neighbor(s). 
As the ripples map to holes, the main interest lies in the size
distribution of distances between clusters and not in the cluster size
distribution itself, as is usually the case. One-dimensional
cluster-cluster aggregation models generally obey
universal dynamical scaling (see~\cite{Hellen:PRE62} and references
therein), which will be seen to be the case also for the sand ripple model.

\section{Noise-induced coarsening ($\gamma < 0$)} \label{admtfsec}

It is known from the mean-field analysis of~\cite{Werner:PRL71}
that for $\gamma = -1$ the average ripple size grows as $\avem(t) \approx
\ln(t+e^{\avem(0)})$. Intuitively, 
the slow growth follows from the fact 
that, for $\gamma < 0$,  
the ripples near extinction are those with the highest
incoming mass rates. Therefore the disappearance of a
ripple involves a rare fluctuation; within the approach
of~\cite{Werner:PRL71}, the mass of a ripple evolving in a
background of mean mass $\avem(t)$ performs a random walk that
is biased away from zero. 

We base our theoretical analysis on this
observation. In what follows, we will assume that at long times, \ie, 
for large $\avem$, the extinction of a ripple is such a rare event,
that it does not affect the ripple size distribution. Neglecting
extinctions, we show that the steady state distribution is given by a product
form. After validating the quasi-static approximation by 
simulations, we use it to show that to leading order in $t$ the average
ripple mass grows as $\avem \sim -\gamma^{-1} \ln (t)$.
Finally we consider the approach to the product measure by studying
nearest neighbor correlations. 

\subsection{Mass distribution} \label{massdistsec}

Without extinctions the ripple size distribution can be solved
exactly. This is due to the short range of interactions between
ripples: The mass transfer rate depends only on the mass at the 
target site. As was noted above in Sec.~\ref{reltootmod}, 
this is similar to the zero-range
process, where the rate depends only on the
site of departure. The most important characteristics of a zero
range process is that its steady state is described by a product
measure~\cite{Evans:BJP}. This was shown to generalize to
processes where the transition rate is a product of functions
of the occupation numbers 
at the site of departure and the target site~\cite{Karimipour:}.
As our model is a special case of this class of models, the
results of~\cite{Karimipour:} apply here as well. For completeness
we give a brief derivation. 

The product measure property implies that the 
stationary probability distribution,
$P(\{m_i\})$, of finding the system in 
configuration $\{m_1,m_2,\ldots,m_N\}$ factorizes as
\begin{equation}
P(\{m_i\}) = \prod_i p(m_i), \label{prodmeasure}
\end{equation}
where $p(m_i)$ is the probability of finding mass $m_i$ at site
$i$. In the steady state there are no correlations between
the ripple sizes.
Starting from the master equation for $P(\{m_i\})$ and using the
product form~\eqref{prodmeasure}, one obtains for the asymmetric case
(the calculation is not presented here since up to index changes it is
identical to that presented in~\cite{Evans:BJP}) 
\begin{equation}
p(m_i)p(m_{i+1})\Gamma(m_i) =
p(m_i+1)p(m_{i+1}-1)\Gamma(m_{i+1}-1). \label{balanceeq} 
\end{equation}

The condition~\eqref{balanceeq}, known as pairwise balance 
\cite{Schutz:JPHYSA}, generalizes
the detailed balance condition familiar from equilibrium
statistical mechanics. It has a simple interpretation. The
left hand size of Eq.~\eqref{balanceeq} represents the mass transfer
to the site $i$ which has to balanced by a transfer out of this site
(the right hand side) in order to be in the steady state. The first
two terms give 
the probability to find a mass $m_i$ at site $i$ with a right
neighbor with mass $m_{i+1}$ and the last term describes the rate at
which the site $i$ gains mass from its neighbor. We emphasize that, 
provided a solution to (\ref{balanceeq}) can be found, this 
proves that the product measure (\ref{prodmeasure}) is an
\emph{exact} stationary solution of the master equation; on the 
basis of general arguments, this solution is then also expected to 
be unique.    

Proceeding similarly for the symmetric dynamics gives (transitions
$\{\ldots,m_{i-1},m_i,m_{i+1},\ldots\} \to
\{\ldots,m_{i-1},m_i+1,m_{i+1}-1,\ldots\}$ and
$\{\ldots,m_{i-1}-1,m_i+1,m_{i+1},\ldots\} \to
\{\ldots,m_{i-1},m_i,m_{i+1},\ldots\}$) 
\begin{eqnarray}
p(m_{i-1})p(m_i)p(m_{i+1})\Gamma(m_i) = \nonumber \\ 
p(m_{i-1}-1)p(m_i+1)p(m_{i+1})\Gamma(m_{i-1}-1).
\end{eqnarray}
Since $p(m_{i+1})$ cancels out we end up with
Eq.~\eqref{balanceeq}. Therefore the steady state distribution is 
independent of the asymmetry of the dynamics. 

Equation~\eqref{balanceeq} can be recast as  
\begin{equation}
\alpha^{-1} := \frac{p(m_i)\Gamma(m_i)}{p(m_i+1)} =
\frac{p(m_{i+1}-1)\Gamma(m_{i+1}-1)}{p(m_{i+1})}, \label{eqfora}
\end{equation}
where $\alpha$ must be a constant. Denoting $p(0) = p_0$ and recursively
iterating equation~\eqref{eqfora}, we obtain
\begin{equation}
p(m) =  p_0 \alpha^m \prod_{i=1}^{m-1} \Gamma(i) 
= p_0 \alpha^m [(m-1)!]^\gamma, \label{exact_eq_for_pk}
\end{equation}
where the product for $m=1$ is defined to give unity and the last form
follows from the definition $\Gamma(m) = m^\gamma$. 

The unknown constants $p_0$ and $\alpha$ can be
determined by the normalization $\sum_{m=0}^\infty p(m) = 1$ and the
expectation value $\avem := \sum_{m=0}^\infty mp(m)$. 
Explicit results for $\gamma = -1$ and $-2$ can be found in 
Appendix~\ref{appdistrcalc}; for $\gamma = -1$, \eqref{exact_eq_for_pk}
is a (shifted) Poisson distribution. In general, the distribution for 
$\avem \gg 1$ can be written as 
\begin{equation}
p(m) 
= C_2(\gamma) e^{\gamma \avem}
\avem^{-\gamma m -(1-\gamma)/2}[(m-1)!]^\gamma, \label{C2fundef}
\end{equation}
where the explicit form of $C_2(\gamma)$ is not important for our
purposes. Using the form given in Eq.~\eqref{C2fundef}, it is
easy to show that the width $\sigma :=
\sqrt{\langle m^2   \rangle - \langle m \rangle^2}$ of the
distribution behaves as $\sigma \sim \sqrt{\avem}$ independent of
$\gamma$.

\begin{figure}
\centering
\includegraphics[width=\linewidth
]{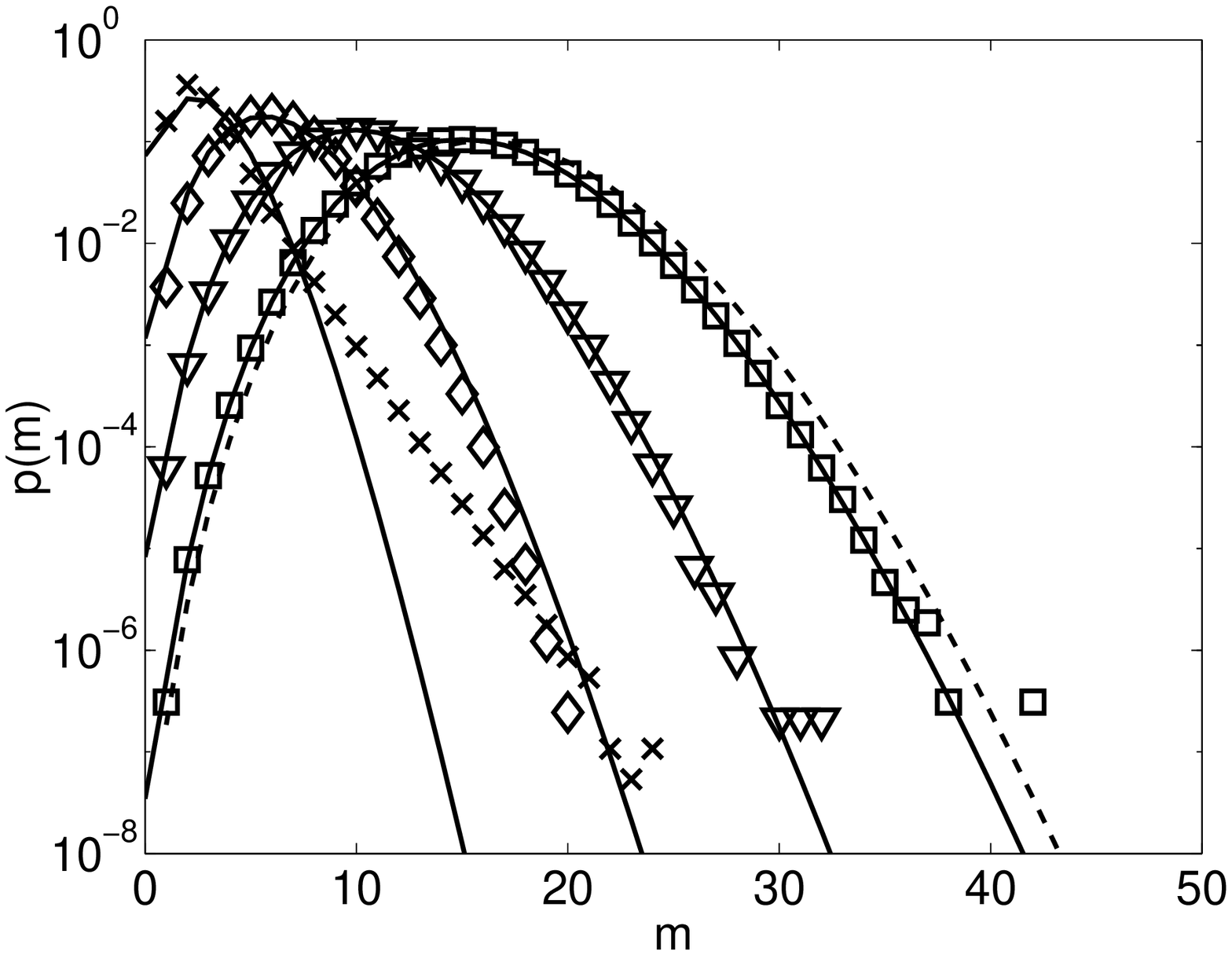}
\caption{The ripple size distributions obtained from simulations for
  $\gamma = -1$ at
  $t = 4~(\times)$, $256~(\diamond)$, $16384~(\nabla)$, and
  $2097152~(\square)$ together with the analytical result (solid lines)
  [Eqs.~\eqref{A3eq} and \eqref{A4eq}]. The initial distribution at $t=0$ is 
  a random one and simulations are averaged over 2000 runs for a
  system of size $M=50000$. The dashed line shows the asymptotic
  solution given by Eq.~\eqref{C2fundef} for $t=2097152$.
  }
\label{pk_sim_vs_theory_gm1}
\end{figure}

\begin{figure}
\centering
\includegraphics[width=\linewidth
]{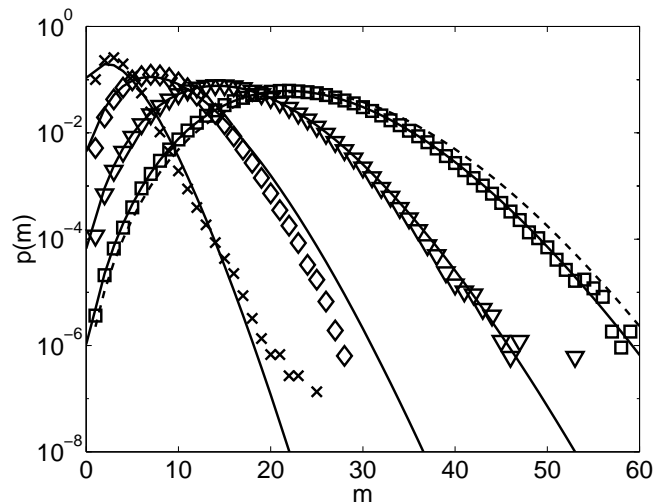}
\caption{The ripples size distributions obtained from simulations for
  $\gamma = -0.5$ at
  $t = 4~(\times)$, $128~(\diamond)$, $4096~(\nabla)$, and
  $131072~(\square)$ together with the analytical result (solid lines)
  [Eq.~\eqref{exact_eq_for_pk}]. The initial distribution at $t=0$ is 
  a random one and simulations are averaged over 500 runs for a
  system of size $M=50000$. The dashed line shows the asymptotic
  solution given by Eq.~\eqref{C2fundef} for $t=131072$.
  }
\label{pk_sim_vs_theory_gm05}
\end{figure}

The calculated distributions are compared to numerics in
Figs.~\ref{pk_sim_vs_theory_gm1} and \ref{pk_sim_vs_theory_gm05}.  
The average ripple mass is not a constant as the simulations include
also ripple extinction. The excellent agreement at long times shows
that indeed these 
become so rare, that between subsequent extinctions
the system has time to equilibrate to the steady state.
Note that all initial distributions converge to the
universal distribution $p(m;t)$ given by Eq.~\eqref{C2fundef}, where
the time-dependence enters only through the mean ripple mass $\avem(t)$.

\subsection{Coarsening law} \label{coarssec}

\begin{figure}
\centering
\includegraphics[angle=-90,width=\linewidth
]{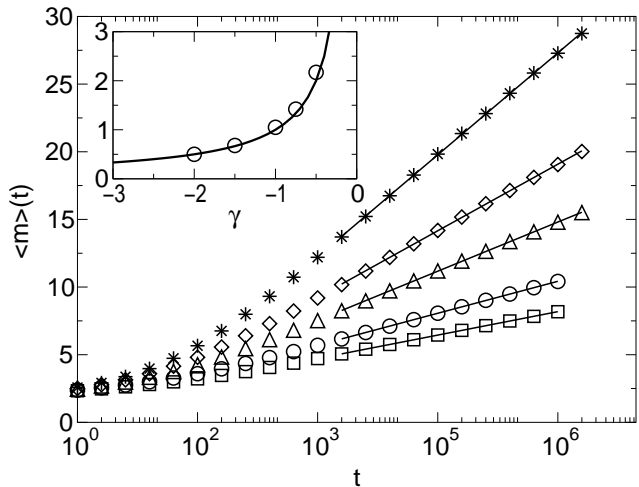}
\caption{The growth of the average ripple mass a function of time for
  $\gamma = -0.5~(*)$, $-0.75~(\diamond)$, $-1~(\Delta)$,
  $-1.5~(\bigcirc)$, and $-2~(\square)$. The least squares fits are shown by
  solid lines. The inset compares the fitted prefactors ($\bigcirc$)
  to the analytic result $-\gamma^{-1}$ (solid line). The
  system sizes range from 50000 to 100000 and averages are taken over at
  least 50 independent runs. 
}
\label{lave_gamma_julkuva2}
\end{figure}

Next we proceed to calculate the 
mean ripple size $\avem(t)$ using an approach 
similar to the analysis of the backgammon model~\cite{Godreche:}.
We assume that, at long times, the probability for a given
ripple to vanish is equal to the probability $p(0)$ obtained
by extrapolating the steady state probability
distribution~\eqref{C2fundef} to $m=0$. The number $N$ of ripples then 
decays according to  $dN/dt \approx -p(0)N$. Since
$\avem(t) = M/N$ we obtain
\begin{equation}
\frac{{\rm d}\avem(t)}{{\rm d}t} \approx p(0) \avem(t) 
\sim e^{\gamma \avem} \avem^{-(1-\gamma)/2},
\end{equation}
which to leading order in $t$ gives
\begin{equation}
\avem(t) \approx  -\gamma^{-1} \ln (t). \label{avemeqatlarget}
\end{equation}
Simulations with different initial conditions are in accord with 
Eq.~\eqref{avemeqatlarget} (Fig.~\ref{lave_gamma_julkuva2}).

\subsection{Decay of correlations} \label{correlationssec}

\begin{figure}
\centering
\includegraphics[width=1.1\linewidth
]{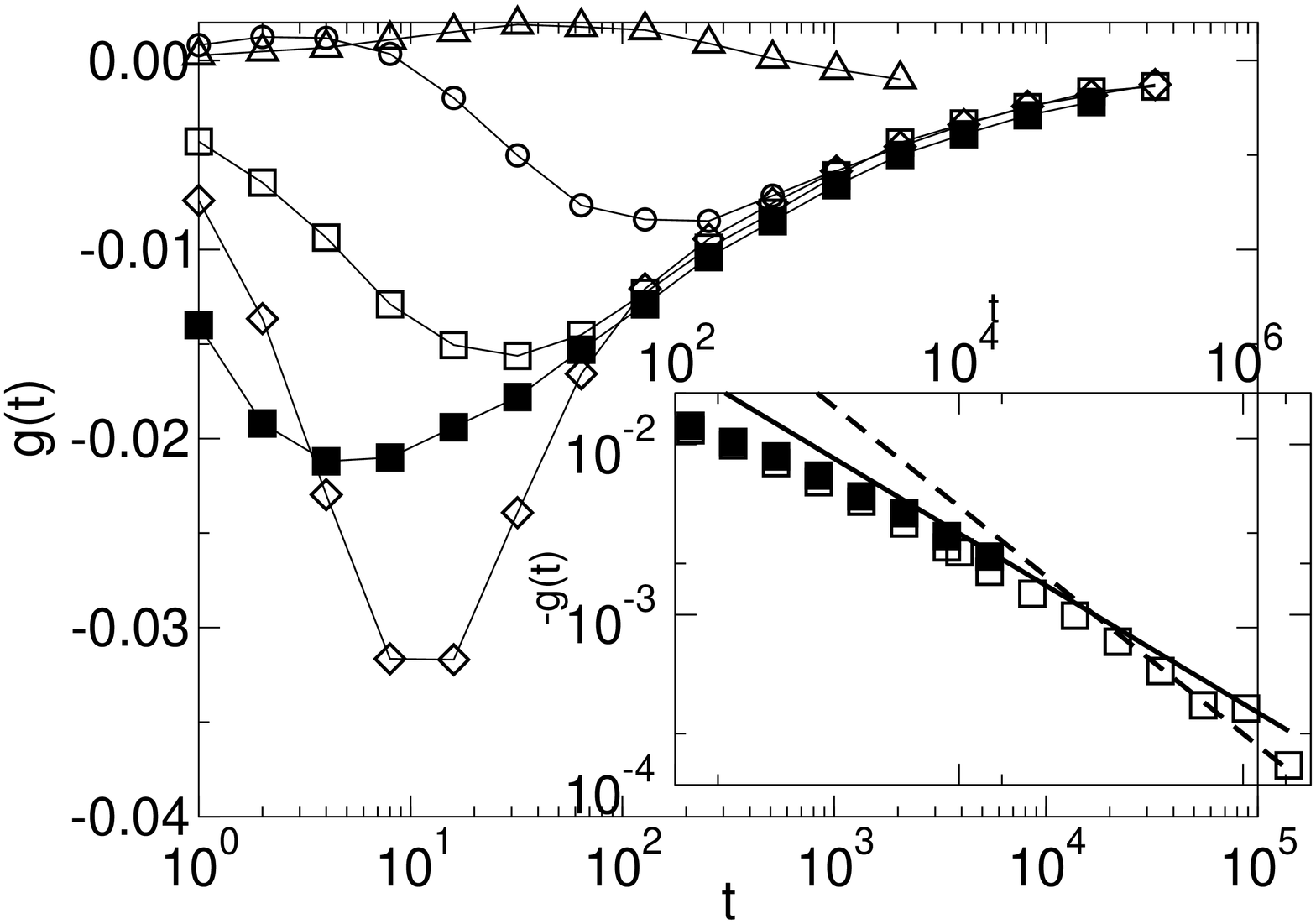}
\caption{The nearest neighbor correlation function $g(t)$ for $\gamma
= -1$. The initial condition is
random~($\avem(0)=2$,$\square$; $\avem(0)=1.2,$\protect\rule{2.2mm}{2.2mm}),
monodisperse~($m(0)=5$,$\diamond$) or Poisson
distributed~($\avem(0)=5$,$\bigcirc$; $\avem(0)=10$,$\Delta$). Open
(filled) symbols correspond to asymmetric (symmetric) dynamics. The
inset shows the decay at late times for the random case. The solid
and dashed lines are guides to the eye with slopes $-1/2$ and $-2/3$,
respectively.  
}
\label{SWORM_gammam1_rand_500_NN3}
\end{figure}

The product measure for the ripple size distribution implies that
there are no correlations between neighboring ripples. This is true
only asymptotically. To study the approach to the product measure
distribution we consider the normalized 
nearest neighbor time correlation
function 
\begin{equation}
g(t) := \frac{\langle m_i
m_{i+1}\rangle - \langle m\rangle^2}{\langle m \rangle^2}.
\end{equation}
As is clear from Fig.~\ref{SWORM_gammam1_rand_500_NN3}, the early
time behavior is 
sensitive to the details of the initial distribution. In this regime
it is possible to have positive correlation between neighboring
ripples but at long times there will always be anticorrelations, \ie,
$g(t) < 0$. The numerically observed correlations seem to be
independent of the initial conditions and vanish in a universal manner as
\begin{equation}
\label{corrdecay}
g(t) \sim -t^{-1/2}
\end{equation}
for both symmetric and asymmetric dynamics.

At first sight one may be tempted to relate the decay of correlations
to the extinction events, which perturb the product measure. However,
as was shown in Sec.~\ref{coarssec}, the probability of extinction
events decays as 
$
p(0) \sim e^{\gamma \avem(t)} \sim t^{-1},
$
which is much faster than the numerically observed decay law
\eqref{corrdecay}. This implies that the power law \eqref{corrdecay}
is associated with the dynamics between extinction events,
which can be described using standard hydrodynamic fluctuation theory
for a one-dimensional system with a single conserved density.

Let $\phi(x,t)$ denote the coarse grained mass fluctuations
in the (quasi-) steady state of mean mass $\avem$. The long wavelength
behavior of $\phi$ is governed by a Langevin equation of the generic
form~\cite{vanBeijeren:PRL54}
\begin{equation}
\label{Burgers}
\frac{\partial \phi}{\partial t} = 
\nu \frac{\partial^2 \phi}{\partial x^2} - \mu \phi 
\frac{\partial \phi}{\partial x} - \frac{\partial \eta}{\partial x},
\end{equation}
where $\eta(x,t)$ is Gaussian white noise with
covariance $\langle \eta(x,t) \eta(x',t') \rangle = 
D \delta(x-x') \delta(t-t')$. For $\mu \neq 0$, Eq.~\eqref{Burgers}
is the noisy Burgers equation~\cite{Forster:PRA}, which has been
widely studied in the context of driven diffusive systems 
\cite{vanBeijeren:PRL54} and interface growth 
\cite{Kardar:PRL56,KrugSpohn_in_Godreches_book,Krug:ADVIP}. 

The coefficients $\nu$, $\mu$, and $D$ appearing in the long wavelength
description can be related to the microscopic dynamics of the sand ripple
model as follows. The nonlinear term on the right hand side of
Eq.~\eqref{Burgers} is generated by the asymmetry, and its coefficient
is given by $\mu = j''(\avem)$, where $j$ is the steady state mass current.
Since in our case $j = \Gamma$, we conclude that 
$\mu \sim \avem^{\gamma - 2}$. In  the symmetric case $\mu = 0$ and
the diffusion coefficient $\nu$ is proportional to  
$\Gamma' \sim \avem^{\gamma - 1}$ (this can be seen 
by expanding Eq.~\eqref{detlambdaeq_sym} around the homogeneous state).
Finally, owing to a fluctuation-dissipation theorem~\cite{Forster:PRA},
the equal time correlations of (\ref{Burgers}) are Gaussian with
covariance $\langle \phi(x) \phi(x') \rangle \sim 
(D/\nu) \delta(x-x')$ independent of $\mu$. As we have
shown above in Sec.~\ref{massdistsec}, in the ripple
model the variance of the mass fluctuations
is always of order $\avem$, hence $D/ \nu \sim  \avem$.   

We want to use equation~\eqref{Burgers} to describe the approach to
the steady state, starting from some initial condition (e.g., the
monodisperse state $\phi \equiv 0$) specified at $t=0$. 
The analysis of Eq.~\eqref{Burgers} shows that at long times,
and for $x \neq x'$,
the pair correlation function takes the scaling
form~\cite{KrugSpohn_in_Godreches_book,Krug:ADVIP} 
\begin{equation}
\label{ScalCorr}
\langle \phi(x,t) \phi(x',t) \rangle = \frac{D}{\nu} 
\frac{1}{\xi(t)} {\cal G}(\xi(t)^{-1}\vert x - x' \vert).
\end{equation}
Here $\cal{G}$ is a scaling function, and $\xi(t)$ denotes
the dynamic correlation length. The prefactor
of the scaling function on the right hand side of Eq.~\eqref{ScalCorr}
is fixed by the requirements that (i) the steady state density fluctuations
are proportional to $D/\nu$, and (ii) the integral over the pair
correlation function is constant due to mass conservation. The correlation
length grows diffusively as $\xi(t) \sim (\nu t)^{1/2}$ for
$\mu = 0$ and superdiffusively as $\xi(t) \sim [(D/\nu)^{1/2} \mu 
t]^{2/3}$ for $\mu \neq 0$.

Keeping $\vert x - x' \vert$ fixed and taking $t \to \infty$,
we see that the pair correlations \eqref{ScalCorr} decay
as $(D/\nu) {\cal G}(0) \xi(t)^{-1}$. 
Expressing $\nu$ and $D/\nu$ in terms of the mean ripple mass,
we conclude that in the symmetric case
($\mu = 0$)  the normalized correlation function 
\eqref{corrdecay} should decay as
\begin{equation}
\label{symdecay}
g(t) \sim \frac{\avem^{-(1 + \gamma)/2}}{t^{1/2}}
\sim \frac{(\ln t)^{-(1+\gamma)/2}}{t^{1/2}}.
\end{equation}
 For $\gamma = -1$ the logarithmic factor disappears and
(\ref{symdecay}) becomes a pure power law with exponent $-1/2$,
in accordance with the simulation results shown in 
Fig.~\ref{SWORM_gammam1_rand_500_NN3}. Moreover,
the explicit calculation for the diffusive case shows that
the scaling function $\cal G$ in Eq.~\eqref{ScalCorr} is negative,
hence $g(t) < 0$ as is observed numerically.

In the asymmetric case the fluctuation theory predicts
an asymptotic decay as $g(t) \sim 1/\xi(t) \sim t^{-2/3}$, with
logarithmic corrections due to the growth of $\avem(t)$. Then why
do we find $g(t) \sim t^{-1/2}$ also for asymmetric dynamics? To 
answer this question, we recall that the asymptotic, superdiffusive
behavior predicted by the noisy Burgers equation 
sets in only beyond a crossover
time scale $t_{\times}$, which increases rapidly with decreasing
strength $\mu$ of the nonlinear term~\cite{Krug:ADVIP}. 
The crossover
time is of the order of $t_\times \sim \nu^5/(D^2 \mu^4)$. 
Inserting the estimates for $\mu$, $\nu$, and
$D/\nu$ derived above, we see that for the ripple model
\begin{equation}
\label{tcross}
t_\times \sim \avem^{3 - \gamma} \sim (\ln t)^{3-\gamma}.
\end{equation}
For the case $\gamma = -1$ considered in 
Fig.~\ref{SWORM_gammam1_rand_500_NN3}, this implies that
superdiffusive behavior can be expected only for times such that
$t/(\ln t)^4 \gg 1$. The left hand side of this inequality becomes
equal to unity for $t \approx 5500$ and reaches the value 10 only
for $t \approx 235000$. Thus the asymptotic regime has not been reached
in our simulations. The slight deviation of the simulation data from
the $t^{-1/2}$-behavior seen after $t = 10^5$ may indicate the
beginning of the crossover.

\section{Unstable coarsening ($\gamma > 0$)} \label{agmtfsec}

For $\gamma>0$ the homogeneous state is linearly unstable
because the largest
ripples are those with the highest growth rate. Ripple
extinction is then no longer a rare event, and the product measure solution
derived in Sec.~\ref{massdistsec} becomes invalid. On the other hand,
it is plausible (and will be confirmed by simulations, see below) 
that the linear
instability supersedes the noise in the time evolution, so that
the deterministic equations~\eqref{detlambdaeq_sym}
and \eqref{detlambdaeq_asym} and the stochastic ripple model show the
same behavior. 

In what follows, we first develop a mean-field theory for the
deterministic model in the simplest case of a linear robber function
($\gamma = 1$). Simulations show that the mean field theory is not
quantitatively correct, presumably due to the neglect of spatial
fluctuations. In the nonlinear regime $0 < \gamma <1$ we use scaling
analysis to derive the coarsening law. 

\subsection{Mean-field analysis for $\gamma=1$} \label{MFgamma1sec}

We start our analysis from the deterministic 
equations~\eqref{detlambdaeq_sym} and \eqref{detlambdaeq_asym}. 
For $\gamma=1$ these become
linear but the system is still non-trivial due to the ripple
extinction. As the system is deterministic, the only randomness lies
in the initial condition. We denote the initial ripple size
distribution by $P_0(\lambda_0)$ and its average by $\bar{\lambda}_0$. 

The mean field approximation consists of replacing the 
ripples surrounding an arbitrary ripple of size $\lambda$ by ripples of the
average size $\avel$, such that the evolution equation becomes
\begin{equation}
\frac{{\rm d}\lambda}{{\rm d}t} = \Gamma(\lambda) -
\Gamma(\avel) = \lambda - \avel.
\label{MFdetlambdaeq}  
\end{equation}
On this level there is no difference between symmetric
and asymmetric mass transfer.
The solution of Eq.~\eqref{MFdetlambdaeq} reads
$\lambda(\lambda_0,t) = e^t[\lambda_0 - F(t)]$, where the function
\begin{equation}
\label{Ft}
F(t) := \int_0^t
{\rm d}\tau \ e^{-\tau} \avel(\tau)
\end{equation} 
has to be calculated
self-consistently. 
Note that at this point we do not explicitly restrict $\lambda(t)$
to be nonnegative (this constraint will enter later).
Once $F(t)$ is known, the ripple size distribution at
time $t$ can be obtained by inverting the
solution for $\lambda(\lambda_0)$ and inserting this into the initial
distribution, with the result 
\begin{equation}
p(\lambda;t) = e^{-t}P_0(e^{-t}\lambda + F(t)). \label{PtMFsol}
\end{equation}
Thus in the mean-field approximation the ripple size
distribution preserves its
initial shape but gets scaled and shifted. 

It is possible to derive a differential equation for the unknown
function $F(t)$. The fraction $\rho(t)$ of surviving ripples is
equal to the probability that $\lambda(t) > 0$, 
\begin{equation} 
\rho(t) = \int_0^\infty {\rm d}\lambda \ p(\lambda;t) =
\int_{F(t)}^\infty {\rm d}x P_0(x) =: P_0^c(F(t)),
\end{equation}
where the last equation defines the cumulative distribution $P_0^c$. 
The average
ripple size is given by $\avel(t) =
\bar{\lambda}_0/\rho(t)$. Inserting this into the definition of $F(t)$ 
and differentiating once gives
\begin{equation}
\frac{{\rm d}F(t)}{{\rm d}t} = \frac{e^{-t}
  \bar{\lambda}_0}{P_0^c(F(t))}. \label{eqforF}
\end{equation}
Hence the problem reduces to solving the
differential equation~\eqref{eqforF} for a given initial distribution
$P_0(\lambda)$.

For example, for an exponential initial distribution $P_0(\lambda_0) =
\bar{\lambda}_0^{-1} e^{-\lambda_0/\bar{\lambda}_0}$, we find 
$F(t) = \bar{\lambda}_0 t$ and
$\avel(t) = \bar{\lambda}_0 e^{t}$, whereas for a flat distribution 
\begin{equation}
P_0(\lambda_0) =
\begin{cases}
(2\bar{\lambda}_0)^{-1}& \lambda_0 \le 2 \bar{\lambda}_0 \\
\ \ \ \ 0& \text{otherwise}
\end{cases}
\end{equation}
the solution is given by $F(t) =
2\bar{\lambda}_0(1-e^{-t/2})$ and $\avel(t) = \bar{\lambda}_0
e^{t/2}$. As the rate of exponential 
growth is different in
these two cases, we conclude that
the coarsening behavior of the mean field
model \eqref{MFdetlambdaeq} is nonuniversal.

In general, the exponential
growth rate of the mean ripple size is governed by the extremal
statistics of the initial distribution $P_0$. 
If the initial ripple sizes are bounded by a maximal size
$\lambda_{\mathrm max}$, and $P_0(\lambda_0) \sim 
(\lambda_{\mathrm{max}} - \lambda_0)^a$ for 
$\lambda_0 \to \lambda_{\mathrm{max}}$, then 
the analysis of Eq.~\eqref{eqforF} shows that
$t^{-1} \ln \avel(t) \to (a + 1)/(a + 2)$, while
for fat initial distributions with a power law tail,
$P_0(\lambda_0) \sim \lambda_0^{-(b + 1)}$, we find
$t^{-1} \ln \avel(t) \to b/(b-1)$.

To compare the predictions of the mean-field theory to simulations
we prefer to show the complement of the cumulative distribution  
\begin{equation}
I(\lambda;t) := \int_{\lambda}^\infty {\rm d}x\ p(x;t) =:  f \left(
\frac{\lambda}{\avel(t)} \right),
\end{equation}
where the last equation defines the scaling function $f(x)$. A similar
definition applies to $p(m;t)$ with the integral replaced by a sum. 
In the case of an exponential distribution $p(\lambda;t)$ also the function
$f(x)$ is exponential whereas for a flat $p(\lambda;t)$ it is linear. 

\begin{figure}
\centering
\includegraphics[width=\linewidth
]{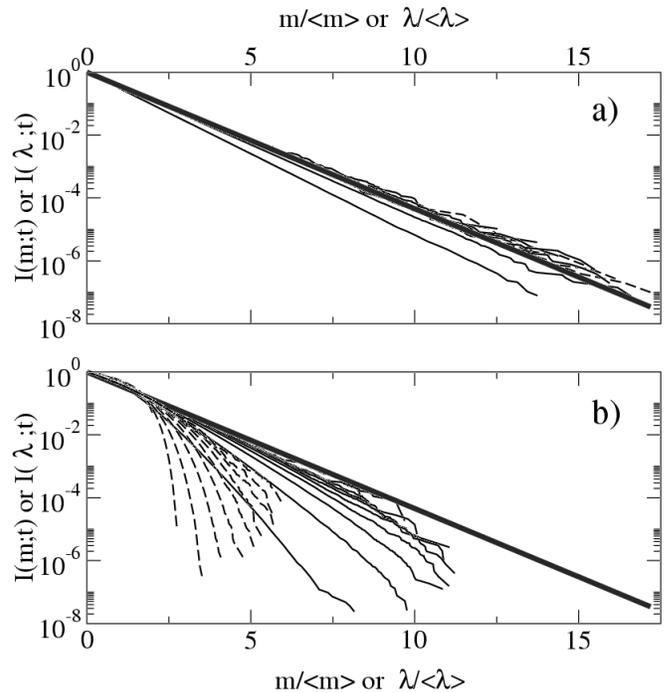}
\caption{The complements of the cumulative ripple size distribution
  for $\gamma=1$. a) The distributions for random (exponential)
  initial distribution with noisy (deterministic) dynamics are denoted
by solid lines (dashed) lines. b) The distributions for monodisperse
(flat) initial distribution with noisy (deterministic) dynamics are denoted
by solid lines (dashed) lines. The curves are shown at
  times $t=1,\ldots,9$ and the thick solid lines in
  both figures represent the function $e^{-m/\avem}$.
}
\label{dists_article_plot2}
\end{figure}

We solved the deterministic equations~\eqref{detlambdaeq_sym} and
\eqref{MFdetlambdaeq} using the 
fourth-order Runge-Kutta method~\cite{NumericalRecipes}. As a check
of the algorithm, we reproduced the solution \eqref{PtMFsol} of
the mean-field equations. For
the full noiseless system~\eqref{detlambdaeq_sym} the exponential initial
distribution remains unchanged [Fig.~\ref{dists_article_plot2}~a);
dashed lines] but
also a flat initial distribution presumably approaches the exponential
one [Fig.~\ref{dists_article_plot2}~b);
dashed lines]. This is in conflict with the
mean-field prediction. In both cases the average ripple size grows as
$\avem(t) \sim e^{t}$. 

Similarly, in the discrete, noisy ripple model the random initial
distribution quickly converges towards an exponential
scaling function~[Fig.~\ref{dists_article_plot2}~a); solid lines]. The
monodisperse 
initial condition spreads out and approaches the same
form~[Fig.~\ref{dists_article_plot2}~b); solid lines]. 
Again, the mean ripple size grows as $\avem(t) \sim
e^{t}$ for both initial distributions. 

Since the deterministic model behaves in a similar manner as the noisy
one, we conclude that,
in contrast to the case $\gamma < 0$, 
the noise is irrelevant. 
The discrepancy between the mean-field theory and the full
deterministic system suggests that the spatial fluctuations are
important, as is often the case for low dimensional systems.
In particular, the numerical results indicate that, in contrast to 
the mean field prediction, the behavior of the full system is universal
with respect to the initial ripple size distribution. 

\subsection{Coarsening law for $0 < \gamma < 1$} \label{gbtw01sec}

As the mean-field equation is not readily solvable for $\gamma \neq 1$
and probably would not describe the problem correctly anyway,
here we present a simple scaling argument 
for the growth of the mean ripple size
in the regime $0 < \gamma < 1$. We start from the observation 
that in the linearly unstable case
($\Gamma'(\lambda) > 0$) predominantly every second ripple grows
and every second one shrinks~\cite{Krug:ACS}. 
Therefore we may consider a simplified system consisting of two
ripples of initial sizes $\lambda_1^0> 
\lambda_2^0$. We calculate the time $t^*$ at which
the average size has doubled. It is given by the conditions
$\lambda_1(t^*) = \lambda_1^0 + \lambda_2^0$ and $\lambda_2(t^*) =
0$. Since the mass is conserved we have $\bar{\lambda} := \lambda_1(t) +
\lambda_2(t) = {\rm const.}$.

Applying Eq.~\eqref{detlambdaeq_sym} gives 
\begin{equation}
\begin{cases} 
\dot{\lambda}_1  & = \lambda_1^\gamma - \lambda_2^\gamma  \\
\dot{\lambda}_2  & = \lambda_2^\gamma - \lambda_1^\gamma,
\end{cases} 
\end{equation}
where the dot denotes derivative with respect to time. 
The solution is implicitly
given by $\lambda_2(t) = \bar{\lambda} - \lambda_1(t)$ and
\begin{equation}
t=\int_{\lambda_1^0}^{\lambda_1(t)} \frac{dx}{x^\gamma -
  (\bar{\lambda}-x)^\gamma}, 
\label{impleq} 
\end{equation}
which together with the definition of $t^*$ implies
the homogeneity relation
\begin{equation}
t^*(a\lambda_1^0,a\bar{\lambda}) = a^{1-\gamma}
t^*(\lambda_1^0,\bar{\lambda}). 
\end{equation}
Assuming that the evolving ripple size distribution is governed
by a single size scale,
it follows that the doubling time depends on the mean
ripple size as $t^* \sim \avel^{1 - \gamma}$. 
The inverse of the doubling time is the growth rate of 
$\avel$. Hence we may write 
\begin{equation}
\frac{{\rm d}\avel(t)}{{\rm d}t} \sim \frac{1}{t^*}\avel(t)
\end{equation}
which yields $\avel(t) \sim t^z$ with $z={1/(1-\gamma)}$. This is
confirmed by simulations, which give $z=1.32 \pm 0.02$ and $1.98 \pm
0.03$ for $\gamma = 0.25$ and $0.50$, respectively. 
We also numerically checked the universal scaling behavior of the
ripple size distribution for $0 < \gamma < 1$,
but in this region the scaling function is more complicated than a
simple exponential.

\section{Conclusions} \label{conlsec}

In this paper we have studied a one-dimensional model for sand ripple
evolution, where
mass is transferred between neighboring sites with algebraic
rates and sites containing no mass are removed from the system (ripple
extinction). 
As the rates depend only on the site to which the mass is 
transferred, the system is similar to a zero range process. Thus the
steady state 
in the absence of ripple extinction is characterized by
a product measure. Asymptotically this continues to 
hold for algebraically decaying mass transfer rates ($\gamma < 0$)
since the extinctions are exponentially rare at late times. As a
consequence the 
average ripple size grows to leading order logarithmically slowly with
the prefactor $-1/\gamma$. 

For $\gamma < 0$ the approach to the steady state product measure is
algebraic. The correlations between masses of neighboring
ripples decay universally, \ie, independent of the initial distribution, as
$t^{-1/2}$ and $t^{-2/3}$ for symmetric and asymmetric mass
transfer, respectively. In the 
asymmetric case the asymptotic regime is preceded by a long crossover where
$t^{-1/2}$-decay is observed. 

For algebraically growing robber functions ($\gamma > 0$) the 
coarsening is driven by the linear instability of the homogeneous state.
Ripple extinctions become frequent, and the product measure is no longer
relevant. The average ripple size grows algebraically as
$t^{1/(1-\gamma)}$ for $0 < \gamma < 1$. The behavior at $\gamma = 0$
is discontinuous since $\avem(t) \sim t^{1/2}$ for $\gamma = 0$, which
follows from the mapping to coalescing random walkers. For $\gamma =1$
$\avem(t) \sim e^t$ and the scaling function of the ripple size
distribution appears to be a simple exponential.  
The dynamical noise, which is necessary to have coarsening for $\gamma
\le 0$, is irrelevant for $\gamma > 0$. 
The mean-field theory developed for $\gamma=1$ 
reproduces the exponential growth of the mean ripple size,
but it is insufficient to
describe the universality of the growth law and the 
ripple size distribution which is observed numerically.

It is interesting to compare the results to the behavior in
one-dimensional cluster-cluster aggregation. Recall that the
model treated here can be mapped to  cluster-cluster aggregation with
hopping rates of clusters depending algebraically on the distance
between them (Sec.~\ref{reltootmod}). When the hopping rates depend as
$\Gamma(m)\sim m^\gamma$ on the {\it masses} of clusters, the growth
of the average cluster 
size is algebraic with $\avem(t) \sim t^{1/(2-\gamma)}$ for all
$\gamma<2$~\cite{Hellen:PRE62}. Thus the behavior for nonnegative
values of $\gamma$ is rather similar in the two models,
but for $\gamma < 0$ one finds a drastic
difference due to the repulsive interaction between clusters in the
ripple model. 

We conclude by adducing some open problems for future studies. One of
the most interesting issues is to understand the coarsening 
and the final ripple size selection in the case of a nonmonotonic 
robber function. This has direct applications in the coarsening of
vortex ripples, where the robber function has recently been
measured~\cite{Andersen:preprint}. Initially these systems are in the
unstable regime, where the transfer function is monotonically
increasing. As we have seen in the present article, even 
this is a harder problem than the case where the dominant
contribution to coarsening comes from the dynamical
fluctuations. For nonmonotonic robber functions one needs an
understanding of both coarsening mechanisms. 
Therefore the starting point into this
direction would be to better understand the $\gamma > 0$ case. 

\section*{Acknowledgements}

E.K.O.H. thanks M.J.~Alava for valuable discussions 
and a critical reading of a preliminary version of the manuscript. He
is also indebted to the kind hospitality of the University of Essen
where the main part of this research was performed. 
J.K. is grateful to K.H.~Andersen for introducing him to the
problem, and to M.R.~Evans for mentioning the backgammon model. 
The financial support by Vilho, Yrj\"o, and Kalle V\"ais\"al\"a
Foundation, DFG within SFB 237, and the Academy of Finland's Center of
Excellence program are gratefully acknowledged. 

\appendix*

\section{Calculation of mass distributions for $\gamma = -1$ and $-2$}
\label{appdistrcalc} 

Here we calculate the explicit form of the mass distribution
\begin{equation}
p(m) = p_0 \alpha^m  \prod_{i=1}^{m-1} \Gamma(i) = p_0
\alpha^{m}[(m-1)!]^\gamma, \label{appeq} 
\end{equation}
for $\Gamma(m) = m^{\gamma}$ in the cases $\gamma =-1$ and $-2$. 
For $m=1$ the
product in 
Eq.~\eqref{appeq} for $m=1$ is defined to be unity, and
we set $(-1)! = 1$.
The normalization condition $\sum_{m=0}^\infty p(m) = 1$ gives
\begin{equation}
p(m) = 
\begin{cases}
(1+\alpha e^\alpha)^{-1} \alpha^m/(m-1)!    & \text{for} \ \ \gamma=-1  \\
I_0^{-1}(2\sqrt{\alpha}) \alpha^{m-1}/[(m-1)!]^2 & \text{for}
\ \ \gamma = -2,
\end{cases} \label{A3eq}
\end{equation}
where $I_n(x)$ is the modified Bessel function of the first kind.
The parameter $\alpha$ is related to the expectation value $\avem =
\sum_{m=0}^\infty mp(m)$ by
\begin{equation}
\avem =
\begin{cases}
\alpha(\alpha+1)e^\alpha/\left(1+\alpha e^\alpha\right)    & \text{for} \ \ \gamma=-1  \\
1+ \sqrt{\alpha}
I_1(2\sqrt{\alpha})/I_0(2\sqrt{\alpha}) & \text{for}\ \ \gamma = -2.
\end{cases} \label{A4eq}
\end{equation}
Using the expansions $I_n(x) = e^{x}/\sqrt{2\pi x} + {\mathcal
  O}(1/x)$ for $\alpha \to \infty$, these formulae simplify to $\avem 
  \approx \alpha$ 
and $\avem \approx \sqrt{\alpha}$, for $\gamma =-1$ and $-2$,
respectively. 
Hence for $\avem \to \infty$ the distributions become
\begin{equation}
p(m) \approx
\begin{cases}
e^{-\avem} \avem^{m-1}/(m-1)!   & \text{for} \ \ \gamma=-1  \\
2\sqrt{\pi} e^{-2\avem}
\avem^{2m-3/2}/[(m-1)!]^2   & \text{for}\ \ \gamma = -2,
\end{cases} \label{exacteqsforg1andg2}
\end{equation}
which are of the general form indicated in Eq.~\eqref{C2fundef}.


\begin{thebibliography}{10}
\bibitem{Bagnold}
R.~A. Bagnold, {\em The Physics of Blown Sand and Desert Dunes} (Chapman \&
  Hall, London, 1941).

\bibitem{Andersen:preprint}
K.~H. Andersen {\it et~al.}, cond-mat/0201529  .

\bibitem{Andersen:PRE63}
K.~H. Andersen, M.-L. Chabanol, and M. {van Hecke}, Phys. Rev. E {\bf 63},
  066308  (2001).

\bibitem{Krug:ACS}
J. Krug, Advances in Complex Systems {\bf 4},  353  (2001).

\bibitem{Werner:PRL71}
B.~T. Werner and D.~T. Gillespie, Phys. Rev. Lett. {\bf 71},  3230  (1993).

\bibitem{Spitzer:AIM5}
F. Spitzer, Adv. in Math. {\bf 5},  246  (1970).

\bibitem{Spohninkirja}
H. Spohn, {\em Large Scale Dynamics of Interacting Particles} (Springer-Verlag,
  New York, 1991).

\bibitem{Evans:BJP}
M.~R. Evans, Brazilian Journal of Physics {\bf 30},  42  (2000),
  cond-mat/0007293.

\bibitem{vanderMeer:PRE63}
D. van~der Meer, K. van~der Weele, and D. Lohse, Phys. Rev. E {\bf 63},  061304
   (2001).

\bibitem{Godreche:}
C. Godr\`eche and J.~M. Luck, to appear in J. Phys. Cond. Matt  (2002),
  cond-mat/0109213.

\bibitem{Ritort:PRL75}
F. Ritort, Phys. Rev. Lett. {\bf 75},  1190  (1995).

\bibitem{Liggetinkirja}
T.~M. Liggett, {\em Interacting Particle Systems} (Springer-Verlag, New York,
  1985).

\bibitem{DerridaEvans_in_Privmans_book}
B. Derrida and M.~R. Evans,  in {\em Nonequilibrium Statistical Mechanics in
  One Dimension}, 1st  ed., edited by V. Privman (Cambridge University Press,
  Cambridge, 1997), pp.\ 277--304.

\bibitem{Majumdar:JSP99}
S.~N. Majumdar, S. Krishnamurthy, and M. Barma, J. Stat. Phys. {\bf 99},  1
  (2000).

\bibitem{ben-Avraham_in_Privmans_book}
D. {ben-Avraham},  in {\em Nonequilibrium Statistical Mechanics in One
  Dimension}, 1st  ed., edited by V. Privman (Cambridge University Press,
  Cambridge, 1997), pp.\ 29--50.

\bibitem{Hellen:PRE62}
E.~K.~O. Hell\'en, T.~P. Simula, and M.~J. Alava, Phys. Rev. E {\bf 62},  4752
  (2000).

\bibitem{Karimipour:}
V. Karimipour and B.~H. Seradjeh, cond-mat/0104196  .

\bibitem{Schutz:JPHYSA}
G. Sch\"utz, R. Ramaswamy, and M. Barma, J. Phys. A {\bf 29},  837  (1996).

\bibitem{vanBeijeren:PRL54}
H. van Beijeren, R. Kutner, and H. Spohn, Phys. Rev. Lett. {\bf 54},  2026
  (1985).

\bibitem{Forster:PRA}
D. Forster, D. Nelson, and M. Stephen, Phys. Rev. A {\bf 16},  732  (1977).

\bibitem{Kardar:PRL56}
M. Kardar, G. Parisi, and Y. Zhang, Phys. Rev. Lett. {\bf 56},  889  (1986).

\bibitem{KrugSpohn_in_Godreches_book}
J. Krug and H. Spohn,  in {\em Solids Far From Equilibrium}, 1st  ed., edited
  by C. Godr\`eche (Cambridge University Press, Cambridge, 1991), pp.\
  479--582.

\bibitem{Krug:ADVIP}
J. Krug, Adv. in Phys. {\bf 46},  139  (1997).

\bibitem{NumericalRecipes}
W.~H. Press, B.~P. Flannery, S.~A. Teukolsky, and W.~T. Vetterling,  in {\em
  Numerical Recipes} (Cambridge University Press, New York, 1986), Chap.~15.1.

\end{thebibliography}
\end{document}